\documentclass[12pt]{article}
\pdfoutput=1
\usepackage[makeroom]{cancel}
\usepackage{amsfonts}
\usepackage{amssymb}
\usepackage{amsmath}
\usepackage{amstext}
\usepackage{graphicx}
\usepackage{epsfig}
\usepackage{verbatim}
\usepackage{fancyhdr}
\usepackage{fancybox}
\usepackage{color}
\usepackage{ulem,bbold}
\usepackage{enumitem}
\usepackage{subfigure}
\usepackage{bbm}
\usepackage{parskip}
\usepackage{latexsym}
\usepackage{theorem}
\usepackage{dsfont}
\usepackage{wasysym}
\usepackage[numbers,sort&compress]{natbib}

\newcommand{\Comment}[1]{{}}
\definecolor{MyDarkBlue}{rgb}{0.15,0.15,0.45}
\usepackage[linktocpage=true]{hyperref}
\hypersetup{
colorlinks=true,
citecolor=MyDarkBlue,
linkcolor=MyDarkBlue,
urlcolor=MyDarkBlue,
}

\newcommand{\bea}{\begin{eqnarray}}  
\newcommand{\eea}{\end{eqnarray}}

\def\nc{\, , \qquad}

\def\be#1\ee{\begin{align}#1\end{align}}

\def\grad{\nabla}
\def\({\left(}
\def\){\right)}
\def\[{\left[}
\def\]{\right]}

\def\dt{\mathrm{d} t \,}

\def\ddx{\mathrm{d}^d x \,}

\def\H{\mathcal{H}}

\def\l{\lambda}

\def\w{\omega}

\def\W{\Omega}
\def\L{\Lambda}

\def\gradt{\tilde{\grad}}
\def\Ht{\tilde{\H}}

\def\Rt{\tilde{R}}

\def\pW{\pi_{\Omega}}
\def\pt{\tilde{\pi}}

\def\th{\tilde{h}}

\begin{document}

\begin{center}
{\Large \bf{Emergent Lorentz Covariance}}
\end{center}

%% JK: Tried a slightly punchier title.

\begin{center}
{\bf{How General Relativity and Lorentz Covariance Arise from the Spatially Covariant Effective Field Theory of the Transverse, Traceless Graviton}}
\end{center}

\vspace{1truecm}

\thispagestyle{empty} \centerline{
{\large  Justin Khoury${}^{a}$, Godfrey E. J. Miller${}^{b}$, and Andrew J. Tolley${}^{c}$}}

\vspace{1cm}

\centerline{{\it ${}^a$
Center for Particle Cosmology, Department of Physics \& Astronomy, University of Pennsylvania}}
\centerline{{\it 209 South 33rd Street, Philadelphia, PA 19104}}

\vspace{1cm}

\centerline{{\it ${}^b$
Princeton Consultants, Inc.}}
\centerline{{\it 2 Research Way, Princeton, NJ 08540}}

\vspace{1cm}

\centerline{{\it ${}^c$
Department of Physics, Case Western Reserve University}}
\centerline{{\it 10900 Euclid Ave, Cleveland, OH 44106}}

\vspace{1cm}

\begin{abstract}
Traditional derivations of general relativity from the graviton degrees of freedom assume space-time Lorentz covariance as an axiom.  In this essay, we survey recent evidence that general relativity is the unique spatially-covariant effective field theory of the transverse, traceless graviton degrees of freedom.  The Lorentz covariance of general relativity, having not been assumed in our analysis, is thus plausibly interpreted as an {\it accidental} or {\it emergent} symmetry of the gravitational sector.  From this point of view, Lorentz covariance is a necessary feature of low-energy graviton dynamics, not a property of space-time.  This result has revolutionary implications for fundamental physics.
\end{abstract}

\vspace{1cm}

\begin{center}
\bf{HONORABLE MENTION}
\end{center}
\begin{center}
\bf{Essay written for the Gravity Research Foundation}
\end{center}
\begin{center}
\bf{2014 Awards for Essays on Gravitation}
\end{center}

%\thispagestyle{empty}
%\end{titlepage}

%%%%%%%%%%%%%%%%%%%%%%%%%%%%%%%%%%%%%%%%%%%%%%%%%%%%%%%%%%%%%%%%%%%%%%%%%%%%%%%%%%%%%%%%%%%%%%%%%%%%%%%%%%%%%%%%%%%%%%%%%%%%%%%%%
\newpage
\setcounter{page}{1}
%\tableofcontents

\noindent{\it \large 1. Introduction}

For over a century, physicists have grappled with the logical foundations of general relativity (GR).  Einstein arrived at GR by combining his gravitational equivalence principle with the Lorentz covariance of special relativity.  Quantization yields a particle, the {\it graviton}, with two transverse, traceless polarizations. The local gravitational degrees of freedom are thus determined by the macroscopic properties of gravitational fields and the assumed space-time symmetries.

Using the formalism of quantum field theory, it is possible to invert this reasoning: GR is the unique Lorentz-covariant theory of a self-interacting massless spin-2 particle, and the equivalence principle results from the form of the interactions in the long-wavelength limit~\cite{Feynman:1996kb,Weinberg:1965rz,Deser:1969wk}. In this picture, GR follows directly from 1)~the local gravitational degrees of freedom and 2)~space-time symmetries.

Without question, GR is the most successful extant paradigm for the interpretation of gravitational phenomena.  Nonetheless, there are compelling reasons to question the basis of gravitational theory.  One motivation is to explain empirical anomalies, most notably the observed magnitude of the cosmic acceleration.  Another is the pursuit of theoretical understanding; by placing GR alongside alternative approaches, we can more clearly perceive the role played by each of its logical elements.

In this essay, we survey recent theoretical progress demonstrating that GR can be derived from the local gravitational degrees of freedom {\it without} assuming space-time symmetry~\cite{Khoury:2011ay,Khoury:2013oqa}.  Our approach relies instead on the weaker assumption of {\it spatial covariance} in the context of the effective field theory of the transverse, traceless gravitons.  By way of phenomenological motivation, the observed cosmic rest frame provides a strong justification for assuming spatial covariance, and the claimed detection of B-modes in the CMB is powerful evidence for the transverse, traceless graviton polarizations~\cite{Ade:2014xna}.  Our result implies that Lorentz covariance is a necessary feature of low-energy graviton dynamics, not a property of space-time: Lorentz covariance is an {\it emergent} symmetry of the gravitational sector.

\vspace{0.5cm}
\noindent{\it \large 2. Relaxing Lorentz Covariance}

In canonical form, GR is a theory of a spatial metric $h_{i j}$ subject to first-class constraints $\H_\mu$.  Each constraint generates a space-time gauge symmetry and eliminates a physical degree of freedom.  The Hamiltonian constraint $\H_0$ generates local time reparameterizations and eliminates the scalar polarization of the graviton; the momentum constraints $\H_i$ generate spatial diffeomorphisms and eliminate the longitudinal polarizations of the graviton.  In $3+1$ dimensions, the spatial metric has six components $h_{i j}$ subject to four constraints $\H_\mu$, so the graviton has $6-4=2$ transverse, traceless polarizations. In this framework, Lorentz covariance arises because the constraints $\H_\mu$ obey the Dirac algebra~\cite{Dirac:Lectures,Barbour:2000qg}, the algebra of the deformations of a space-like hypersurface embedded in a Lorentzian space-time manifold~\cite{Teitelboim:1972vw}.

The technical role of the constraints is clarified by a conformal decomposition of the spatial metric.  Following~\cite{Farkas:2010dw}, in $d$ spatial dimensions we define
\be
\th_{i j} \equiv e^{-\W} h_{i j}
\nc
\W \equiv \frac{1}{d} \log h
\, .
\label{thphi}
\ee
By definition, the metric $\th_{i j}$ has unit determinant.  The conformal factor $\W$ controls the local geometric scale, while $\th_{i j}$ defines spatial geometry up to local changes of scale.  As elaborated in~\cite{Khoury:2011ay,Khoury:2013oqa}, the Hamiltonian constraint $\H_0$ removes the scalar polarization of the graviton by rendering the conformal factor $\W$ non-dynamical.  Meanwhile, the momentum constraints $\H_i$ remove the longitudinal polarizations from the unit-determinant metric $\th_{i j}$.

To relax the assumption of space-time Lorentz covariance, we drop the Hamiltonian constraint $\H_0$; to avoid a scalar graviton polarization, we drop the conformal factor $\W$ as a dynamical field.  Having one fewer component to begin with, a spatially-covariant unit-determinant metric $\th_{i j}$ has the same number of degrees of freedom as a space-time covariant metric $h_{i j}$.  In $3+1$ dimensions,
\be
5 \cdot \th_{i j} \, 's
- 3 \cdot \Ht_i  \, 's
= 2 \ {\rm Degrees \ of \ Freedom} \, .
\ee
Here $\Ht_i$ represents the original momentum constraints $\H_i$ after their form has been adapted to the absence of $\W$ as a dynamical field.

By construction, spatially-covariant theories of a unit-determinant metric describe the same transverse, traceless graviton polarizations as GR.  Though the kinematical state space is essentially the same, in principle the dynamical evolution of the graviton degrees of freedom could differ dramatically.  However, as we will see, demanding a consistent algebra and evolution for the momentum constraints singles out GR as the unique possibility.

\vspace{0.5cm}
\noindent{\it \large 3. Dynamical Evolution and Consistency}

To describe the dynamical evolution of $\th_{i j}$, we will construct the most general action consistent with the assumption of spatial covariance.  The canonical action functional of a physical system is defined on paths through phase space. The momentum conjugate to an ordinary spatial metric $h_{i j}$ is a symmetric tensor $\pi^{i j}$. The conformal factor $\W$ is conjugate to the trace of $\pi^{i j}$, while $\pt^{i j}$ is proportional to the traceless part of $\pi^{i j}$~\cite{Khoury:2011ay,Khoury:2013oqa}.  Having discarded $\W$ as a dynamical field, our phase space consists of the unit-determinant metric $\th_{i j}$ and the traceless conjugate momentum tensor $\pt^{i j}$. To realize spatial covariance, we impose the momentum constraints $\Ht_i$ by introducing Lagrange multipliers $N^i$ into the canonical action.

In $d+1$ dimensions, the most general spatially-covariant canonical action is
\be
S = \int \dt \ddx \( \dot{\th}_{i j} \pt^{i j} - \pi_H - N^i \Ht_i \)
,
\label{SCGactionintro}
\ee
which includes GR in spatially-covariant gauge as a special case~\cite{Khoury:2011ay,Khoury:2013oqa}.  The scalar function $\pi_H$ is the physical Hamiltonian density; the absence of explicit time-reparameterization invariance on the reduced phase space implies that $\pi_H$ need not vanish on-shell.  Spatial covariance is enforced by the $\Ht_i$ momentum constraints, whose general form~\cite{Khoury:2011ay,Khoury:2013oqa} includes a second scalar function $\pi_K$:
\be
\Ht_i \equiv - 2 \th_{i j} \gradt_k \pt^{j k} - \gradt_i \pi_K
\, ,
\label{kconformintro}
\ee
where $\tilde{\nabla}$ denotes the covariant derivative with respect to $\th_{i j}$.  If $\W$ were dynamical, $\pi_H$ and $\pi_K$ would be fixed by the trace of $\pi^{i j}$; since $\W$ is assumed non-dynamical, $\pi_H$ and $\pi_K$ are arbitrary scalar functions.  By construction, the graviton of this theory lacks a scalar polarization.  Meanwhile, the constraints $\Ht_i$ fix the divergence of the momentum tensor, $\gradt_j \pt^{i j}$, eliminating the longitudinal polarizations.  It follows that the graviton of our theory possesses only the desired transverse, traceless polarizations.

The scalar functions $\pi_H$ and $\pi_K$ may depend on time $t$, the phase space variables $\th_{i j}, \pt^{i j}$, and their spatial derivatives. The leading scalar operator containing spatial derivatives is the Ricci scalar $\Rt \equiv \th^{i k} \th^{j \ell} \Rt_{ijk\ell}$, and it is consistent to impose a cutoff which excludes sub-leading operators~\cite{Khoury:2013oqa}. In the low-energy effective theory, $\pi_H$ and $\pi_K$ depend on spatial gradients exclusively through $\Rt$.  In this formalism, GR corresponds to:
\be
\pi_H^{\rm GR} = - \dot{\W}(t) \pW
\nc
\pi_K^{\rm GR} = \frac{2}{d}  \pW
\, ,
\label{GRHKintro}
\ee
where
\be
\pW \equiv \pm \sqrt{d (d-1)} \sqrt{\tilde{h}_{ik}\tilde{h}_{j \ell} \tilde{\pi}^{ij}\tilde{\pi}^{k\ell} - \Rt \, e^{(d-1)\W(t) } + 2 \L e^{d\W(t)} }
\, .
\label{GRPWintro}
\ee
Here $\Lambda$ is the cosmological constant, and $\Omega(t)$ is a monotonic function of time.

In general, the allowed form for $\pi_H$ and $\pi_K$ is constrained by two considerations:

\noindent $i)$ To generate a consistent gauge symmetry, the momentum constraints $\Ht_i$ must be first-class under the action of the Poisson bracket (with respect to $\th_{i j}, \pt^{i j}$),
\be
\{ \Ht_i(x) , \Ht_a(y) \} \sim 0
\, ,
\label{algebraconintro}
\ee
where $\sim$ denotes {\it weak equality}, {\it i.e.}, equality after the imposition of all constraints~\cite{Henneaux:1992ig,Hanson:1976cn}. The condition~\eqref{algebraconintro} means that the algebra of the constraints is {\it closed}, in the sense that the Poisson bracket of any two constraints is proportional to the constraints. 

\noindent $ii)$ The momentum constraints $\Ht_i$ must be preserved under time evolution:
\be
\dot{\Ht}_i \sim 0
\, .
\ee
These two conditions are remarkably restrictive~\cite{Khoury:2011ay,Khoury:2013oqa}. After some lengthy algebra, we showed that they fix the allowed form for $\pi_H$ and $\pi_K$.  The result is\footnote{Technically, there is a second possibility, namely that $\pi_K$ and $\pi_H$ are ultralocal functions, {\it i.e.}, they depend on $t$ and $\th_{i j}, \pt^{i j}$, but {\it not} on their spatial gradients. This possibility, while not as physically interesting, is a proof of principle that our formalism allows more general theories than GR, albeit in the ultra-local limit.} 
\be
\pi_H = - \dot{\w}(t)\pi_\w
\nc
\pi_K = \frac{2}{d} \pi_\w
\, ,
\ee
where
\be
\pi_\w  \equiv \pm \sqrt{d (d-1)} \sqrt{\phi(2) \pm \Rt \, e^{(d-1)\w(t) } + 2 \l e^{d\w(t)}}\,,
\ee
with $\w(t)$ an arbitrary function of time. 
With the trivial relabeling $\omega \rightarrow \Omega$, $\lambda \rightarrow \Lambda$, this becomes identical to the GR result given by~\eqref{GRHKintro} and~\eqref{GRPWintro}.  
To summarize, the combined requirements of 1) closure of the constraints and 2) consistency under time evolution are sufficient to single out GR as the unique effective field theory of the graviton degrees of freedom.

\vspace{0.5cm}
\noindent {\it \large 4. Conclusion}

We have derived GR from the local gravitational degrees of freedom assuming only spatial covariance.  To our knowledge, this represents an enormous advance over all previous derivations of GR from the graviton degrees of freedom, which assume Lorentz covariance at the outset~\cite{Feynman:1996kb,Weinberg:1965rz,Deser:1969wk}. Our approach relies on the weaker assumption of spatial covariance, and yet achieves an equally powerful result.

In our derivation, consistency forces the operators $\tilde{h}_{ik}\tilde{h}_{j \ell} \tilde{\pi}^{ij}\tilde{\pi}^{k\ell}$ and $\Rt$ to appear in a Lorentz covariant combination, and our cutoff excludes all higher-order operators which might spoil the symmetry.  Both the conformal scale factor $\W(t)$ and the cosmological constant $\L$ arise in the theory as constants of integration.

In light of our proof, it is plausible to interpret Lorentz symmetry in the gravitational sector as an {\it accidental} or {\it emergent} symmetry.  Accidental symmetries arise in an effective field theory when all the allowable operators which violate the symmetry are confined above the energy cutoff; in this respect, it is the opposite of spontaneous symmetry breaking.  From this point of view, Lorentz covariance is a feature of low-energy graviton dynamics, not a property of space-time.

This result has clear implications for fundamental physics.  For generations, Lorentz covariance has been a central pillar in the construction of physical theories.  In this work, gravitational Lorentz covariance has itself been derived from theoretical first principles within the context of a well-defined physical theory.

%{\bf Acknowledgments:} We thank Julian Barbour, Jolyon Bloomfield, Stanley Deser, Tim Koslowski, and C.~Jess Riedel for helpful discussions. G.E.J.M. was supported in part by the Department of Energy under contract No. DE-AC02-76-ER-03071. J.K. was supported in part by NASA ATP grant NNX11AI95G and the Alfred P. Sloan Foundation.  A.J.T. was supported in part by the Department of Energy under grant DE-FG02-12ER41810.


\begin{thebibliography}{99}

%\cite{Feynman:1996kb}
\bibitem{Feynman:1996kb}
R.~P.~Feynman, F.~B.~Morinigo, W.~G.~Wagner and B.~Hatfield,
``Feynman Lectures on Gravitation,''
%\href{http://www.slac.stanford.edu/spires/find/hep/www?ir${\mathcal{N}}\!=3$475700}{SPIRES entry}
{\it  Reading, USA: Addison-Wesley (1995) 232 p. (The advanced book program)}

%\cite{Weinberg:1965rz}
\bibitem{Weinberg:1965rz}
S.~Weinberg,
``Photons and Gravitons in Perturbation Theory: Derivation of Maxwell's and   Einstein's Equations,''
Phys.\ Rev.\  {\bf 138} (1965) B988.
%%CITATION = PHRVA,138,B988;%%

%\cite{Deser:1969wk}
\bibitem{Deser:1969wk}
S.~Deser,
``Self-Interaction and Gauge Invariance,''
Gen.\ Rel.\ Grav.\  {\bf 1} (1970) 9
[arXiv:gr-qc/0411023].
%%CITATION = GRGVA,1,9;%%

%\cite{Khoury:2011ay}
\bibitem{Khoury:2011ay} 
  J.~Khoury, G.~E.~J.~Miller and A.~J.~Tolley,
  ``Spatially Covariant Theories of a Transverse, Traceless Graviton, Part I: Formalism,''
  Phys.\ Rev.\ D {\bf 85}, 084002 (2012)
  [arXiv:1108.1397 [hep-th]].
  %%CITATION = ARXIV:1108.1397;%%

%\cite{Khoury:2013oqa}
\bibitem{Khoury:2013oqa} 
  J.~Khoury, G.~E.~J.~Miller and A.~J.~Tolley,
  ``On the Origin of Gravitational Lorentz Covariance,''
  arXiv:1305.0822 [hep-th].
  %%CITATION = ARXIV:1305.0822;%%
  %1 citations counted in INSPIRE as of 07 Mar 2014
	
%\cite{Ade:2014xna}
\bibitem{Ade:2014xna} 
  P.~A.~R.~Ade {\it et al.}  [BICEP2 Collaboration],
  ``BICEP2 I: Detection Of B-mode Polarization at Degree Angular Scales,''
  arXiv:1403.3985 [astro-ph.CO];
  ``BICEP2 II: Experiment and Three-Year Data Set,''
  arXiv:1403.4302 [astro-ph.CO].
  %%CITATION = ARXIV:1403.4302;%%

%\cite{Dirac:Lectures}
\bibitem{Dirac:Lectures}
P.~A.~M.~Dirac,
``Lectures on Quantum Mechanics,''
{\it Mineola, NY, USA: Dover Publications, Inc. (2001)}
{\it Originally Published in New York, USA by the Belfer Graduate School of Science, Yeshiva Univ. (1964)}.
%P.~A.~M.~Dirac,
%``The Theory of Gravitation in Hamiltonian Form,''
%Proc.\ Roy.\ Soc.\ Lond.\  A {\bf 246} (1958) 333;
%%%CITATION = PRSLA,A246,333;%%
%P.~A.~M.~Dirac,
%``Generalized Hamiltonian Dynamics,''
%Proc.\ Roy.\ Soc.\ Lond.\  A {\bf 246} (1958) 326;
%%%CITATION = PRSLA,A246,326;%%
%P.~A.~M.~Dirac,
%``The Hamiltonian Form of Field Dynamics,''
%Can.\ J.\ Math.\  {\bf 3} (1951) 1;
%%%CITATION = CJMAA,3,1;%%
%P.~A.~M.~Dirac,
%``Generalized Hamiltonian Dynamics,''
%Can.\ J.\ Math.\  {\bf 2} (1950) 129.
%%%CITATION = CJMAA,2,129;%%
%%\cite{Schwinger:1963zza}
%%\bibitem{Schwinger:1963zza} 
%  J.~Schwinger,
%  ``Commutation Relations and Conservation Laws,''
%  Phys.\ Rev.\  {\bf 130}, 406 (1963).
%  %%CITATION = PHRVA,130,406;%%
%%\cite{Schwinger:1963zz}
%%\bibitem{Schwinger:1963zz} 
%  J.~Schwinger,
%  ``Energy and Momentum Density in Field Theory,''
%  Phys.\ Rev.\  {\bf 130}, 800 (1963).
%  %%CITATION = PHRVA,130,800;%%
%  %75 citations counted in INSPIRE as of 07 Nov 2013

%\cite{Barbour:2000qg}
\bibitem{Barbour:2000qg}
J.~Barbour, B.~Z.~Foster and N.~O'Murchadha,
``Relativity without Relativity,''
Class.\ Quant.\ Grav.\  {\bf 19} (2002) 3217
[arXiv:gr-qc/0012089].
%%CITATION = CQGRD,19,3217;%%

%\cite{Teitelboim:1972vw}
\bibitem{Teitelboim:1972vw}
C.~Teitelboim,
``How Commutators of Constraints Reflect the Space-Time Structure,''
Annals Phys.\  {\bf 79} (1973) 542.
%%CITATION = APNYA,79,542;%%

%\cite{Farkas:2010dw}
\bibitem{Farkas:2010dw}
S.~Farkas and E.~J.~Martinec,
``Gravity from the Extension of Spatial Diffeomorphisms,''
arXiv:1002.4449 [hep-th].
%%CITATION = ARXIV:1002.4449;%%

%%\cite{Barbour:2011dn}
%\bibitem{Barbour:2011dn}
%J.~Barbour,
%``Shape Dynamics. an Introduction,''
%arXiv:1105.0183 [gr-qc];
%%%CITATION = ARXIV:1105.0183;%%
%J.~Barbour and N.~O.~Murchadha,
%``Conformal Superspace: the Configuration Space of General Relativity,''
%arXiv:1009.3559 [gr-qc];
%%%CITATION = ARXIV:1009.3559;%%
%E.~Anderson, J.~Barbour, B.~Z.~Foster, B.~Kelleher and N.~O.~Murchadha,
%``The Physical Gravitational Degrees of Freedom,''
%Class.\ Quant.\ Grav.\  {\bf 22} (2005) 1795
%[arXiv:gr-qc/0407104];
%%%CITATION = CQGRD,22,1795;%%
%E.~Anderson, J.~Barbour, B.~Z.~Foster, B.~Kelleher and N.~O'Murchadha,
%``A First-Principles Derivation of York Scaling and the Lichnerowicz-York   Equation,''
%arXiv:gr-qc/0404099;
%%%CITATION = GR-QC/0404099;%%
%E.~Anderson, J.~Barbour, B.~Foster and N.~O'Murchadha,
%``Scale-Invariant Gravity: Geometrodynamics,''
%Class.\ Quant.\ Grav.\  {\bf 20} (2003) 1571
%[arXiv:gr-qc/0211022].
%%%CITATION = CQGRD,20,1571;%%

%%\cite{Gomes:2011zj}
%\bibitem{Gomes:2011zj}
%H.~Gomes,
%``The Coupling of Shape Dynamics to Matter,''
%arXiv:1112.0374 [gr-qc];
%%%CITATION = ARXIV:1112.0374;%%
%H.~Gomes and T.~Koslowski,
%``Coupling Shape Dynamics to Matter Gives space-time,''
%arXiv:1110.3837 [gr-qc];
%%%CITATION = ARXIV:1110.3837;%%
%T.~Koslowski,
%``Shape Dynamics,''
%arXiv:1108.5224 [gr-qc];
%%%CITATION = ARXIV:1108.5224;%%
%T.~Budd and T.~Koslowski,
%``Shape Dynamics in 2+1 Dimensions,''
%arXiv:1107.1287 [gr-qc];
%%%CITATION = ARXIV:1107.1287;%%
%H.~Gomes, S.~Gryb, T.~Koslowski and F.~Mercati,
%``The Gravity/CFT Correspondence,''
%arXiv:1105.0938 [gr-qc];
%%%CITATION = ARXIV:1105.0938;%%
%H.~Gomes and T.~Koslowski,
%``The Link Between General Relativity and Shape Dynamics,''
%arXiv:1101.5974 [gr-qc];
%%%CITATION = ARXIV:1101.5974;%%
%H.~Gomes, S.~Gryb and T.~Koslowski,
%``Einstein Gravity as a 3D Conformally Invariant Theory,''
%Class.\ Quant.\ Grav.\  {\bf 28} (2011) 045005
%[arXiv:1010.2481 [gr-qc]].
%%%CITATION = CQGRD,28,045005;%%

%%\cite{Horava:2010zj}
%\bibitem{Horava:2010zj}
%P.~Ho\v{r}ava and C.~M.~Melby-Thompson,
%``General Covariance in Quantum Gravity at a Lifshitz Point,''
%Phys.\ Rev.\  D {\bf 82} (2010) 064027
%[arXiv:1007.2410 [hep-th]].
%%%CITATION = PHRVA,D82,064027;%%

%%\cite{Gourgoulhon:2007nr}
%\bibitem{Gourgoulhon:2007nr} 
%E.~Gourgoulhon,
%``3+1 formalism and bases of numerical relativity,''
%arXiv:gr-qc/0703035.
%  %%CITATION = GR-QC/0703035;%%
%  %121 citations counted in INSPIRE as of 09 Feb 2014

%%\cite{Henneaux:1989zc}
%\bibitem{Henneaux:1989zc} 
%  M.~Henneaux and C.~Teitelboim,
%  ``The Cosmological Constant And General Covariance,''
%  Phys.\ Lett.\ B {\bf 222}, 195 (1989).
%  %%CITATION = PHLTA,B222,195;%%
%  %103 citations counted in INSPIRE as of 13 Feb 2013

%%\cite{Arnowitt:1962hi}
%\bibitem{Arnowitt:1962hi}
%R.~L.~Arnowitt, S.~Deser and C.~W.~Misner,
%``The Dynamics of General Relativity,''
%arXiv:gr-qc/0405109.
%%%CITATION = GR-QC/0405109;%%

%\cite{Henneaux:1992ig}
\bibitem{Henneaux:1992ig}
M.~Henneaux and C.~Teitelboim,
``Quantization of Gauge Systems,''
%\href{http://www.slac.stanford.edu/spires/find/hep/www?ir${\mathcal{N}}\!=2$824396}{SPIRES entry}
{\it  Princeton, USA: Univ. Pr. (1992) 520 p}

%\cite{Hanson:1976cn}
\bibitem{Hanson:1976cn} 
  A.~J.~Hanson, T.~Regge and C.~Teitelboim,
  ``Constrained Hamiltonian Systems,''
  RX-748.
  %%CITATION = RX-748;%%
  
%%\cite{Belinsky:1970ew}
%\bibitem{Belinsky:1970ew}
%V.~A.~Belinsky, I.~M.~Khalatnikov and E.~M.~Lifshitz,
%``Oscillatory Approach to a Singular Point in the Relativistic Cosmology,''
%Adv.\ Phys.\  {\bf 19} (1970) 525.
%%%CITATION = ADPHA,19,525;%%

%%\cite{Henneaux:2007ej}
%\bibitem{Henneaux:2007ej}
%M.~Henneaux, D.~Persson and P.~Spindel,
%``Spacelike Singularities and Hidden Symmetries of Gravity,''
%Living Rev.\ Rel.\  {\bf 11} (2008) 1
%[arXiv:0710.1818 [hep-th]];
%%%CITATION = 00222,11,1;%%
%T.~Damour, M.~Henneaux and H.~Nicolai,
%``Cosmological Billiards,''
%Class.\ Quant.\ Grav.\  {\bf 20} (2003) R145
%[arXiv:hep-th/0212256].
%%%CITATION = CQGRD,20,R145;%%

%%\cite{Salopek:1990mp}
%\bibitem{Salopek:1990mp}
%D.~S.~Salopek,
%``Nonlinear Solutions of Long Wavelength Gravitational Radiation,''
%Phys.\ Rev.\  D {\bf 43} (1991) 3214;
%%%CITATION = PHRVA,D43,3214;%%
%D.~S.~Salopek and J.~R.~Bond,
%``Stochastic Inflation and Nonlinear Gravity,''
%Phys.\ Rev.\  D {\bf 43} (1991) 1005;
%%%CITATION = PHRVA,D43,1005;%%
%D.~S.~Salopek and J.~R.~Bond,
%``Nonlinear Evolution of Long Wavelength Metric Fluctuations in Inflationary   Models,''
%Phys.\ Rev.\  D {\bf 42} (1990) 3936.
%%%CITATION = PHRVA,D42,3936;%%

%%\cite{Tolley:2008na}
%\bibitem{Tolley:2008na}
%A.~J.~Tolley and M.~Wyman,
%``Stochastic Inflation Revisited: Non-Slow Roll Statistics and Dbi Inflation,''
%JCAP {\bf 0804} (2008) 028
%[arXiv:0801.1854 [hep-th]].
%%%CITATION = JCAPA,0804,028;%%

%%\cite{bloomfield}
%\bibitem{bloomfield}
%J.~Bloomfield and G.~E.~J.~Miller, to appear.

%%\cite{Bellorin:2013zbp}
%\bibitem{Bellorin:2013zbp} 
%  J.~Bellorin, A.~Restuccia and A.~Sotomayor,
%  ``A consistent Horava gravity without extra modes and equivalent to general relativity at the linearized level,''
%  Phys.\ Rev.\ D {\bf 87}, 084020 (2013)
%  [arXiv:1302.1357 [hep-th]].
%  %%CITATION = ARXIV:1302.1357;%%

%%\cite{Weinberg:1972gc}
%\bibitem{Weinberg:1972gc}
%S.~Weinberg,
%``Gravitation and Cosmology: Principles and Applications of the General Theory of Relativity,''
%{\it  New York, NY, USA: John Wiley \& Sons, Inc. (1972) 657 p.}



\end{thebibliography}
\end{document}